\documentclass{ws-procs9x6}

\begin{document}

\title{The Fermion Sign Problem and High Density Effective Theory}

\author{D.~K.~Hong\footnote{\uppercase{W}ork partially
supported by the academic research fund of the
\uppercase{M}inistry of \uppercase{E}ducation,
\uppercase{R}epublic of \uppercase{K}orea, \uppercase{P}roject
\uppercase{KRF-2000-015-DP0069}, and the \uppercase{USA-K}orea
\uppercase{C}ooperative \uppercase{S}cience \uppercase{P}rogram,
\uppercase{NSF} 9982164.}}

\address{Department of Physics, \\
Pusan National University, \\
Pusan 609-735, Korea\\
E-mail: dkhong@pusan.ac.kr}

\author{S.~D.~H.~Hsu\footnote{\uppercase{W}ork partially
supported by the \uppercase{U.S. D}eptartment of
\uppercase{E}nergy, \uppercase{DE-FG06-85ER40224} and the
\uppercase{USA-K}orea \uppercase{C}ooperative \uppercase{S}cience
\uppercase{P}rogram, \uppercase{NSF} 9982164.}}

\address{Institute of Theoretical Science \\
University of Oregon, \\
Eugene, OR 97403-5203\\
E-mail: hsu@duende.uoregon.edu}

\maketitle

\abstracts{We investigate the positivity of the Euclidean path
integral measure for low-energy modes in dense fermionic matter.
We show that the sign problem usually associated with fermions is
absent if one considers only low-energy degrees of freedom. We
describe a method for simulating dense QCD on the lattice and give
a proof using rigorous inequalities that the color-flavor locked
(CFL) phase is the true vacuum of three flavor, massless QCD.}

\section{Introduction}
Euclidean quantum chromodynamics (QCD) with a non-zero chemical
potential has a complex measure, which has made lattice simulation
particularly difficult\cite{Hands:2001jn}. (Lattice simulation
of the QCD phase boundary at finite
density\cite{Fodor:2001au,Allton:2002zi} has been a topic of
recent interest.) This problem is often referred to as the sign
problem, because, by appropriately grouping terms, quantities such
as the partition function can be written as a sum over real, but
potentially negative, terms. (That this grouping can be
accomplished is in many systems a consequence of a discrete
symmetry such as parity or time-reversal invariance.)
Indeterminate signs are enough to preclude use of importance
sampling, the main technique for speeding up Monte Carlo
integration. It is important to note that while the sign problem
often arises in systems of fermions, it is neither inevitable nor
inescapable. For example, in QCD at zero chemical potential and in
the Hubbard model at half filling one can organize the sum so that
terms are real and positive.

Analytical work in color superconductivity\cite{csc} has
demonstrated a rich phase structure at high density, and
stimulated interest in QCD at non-zero baryon density. Several
experiments have been proposed to probe matter at density of a few
times nuclear matter density\cite{exper}. Even rudimentary
information about the behavior of dense matter would be useful to
the experimental program, as well as to the study of compact
astrophysical objects such as neutron stars.
Recently\cite{Hong:2002nn}, we showed that QCD near a Fermi
surface has positive, semi-definite measure. In the limit of low
energies, the contribution of the remaining modes far from the
Fermi surface can be systematically expanded, using a high density
effective theory previously introduced by one of
us\cite{Hong:2000tn,Hong:2000ru}. This effective theory is
sufficient to study phenomena like color superconductivity,
although quantities like the equation of state may be largely
determined by dynamics deep in the Fermi sea.

The expansion about the Fermi surface is in powers of $1/ \mu$,
where $\mu$ is the chemical potential. For this expansion to be
controlled, the ultraviolet cutoff of our effective theory must be
less than $\mu$, or equivalently the scale of the physics of
interest must be small relative to the chemical potential. In QCD
at asymptotic density, the superconducting gap is exponentially
small, so this condition is satisfied. However, it is also quite
possible that at intermediate densities (e.g., those inside a
neutron star) the gap is somewhat smaller than $\mu$, providing us
with an additional small dimensionless parameter. Even if this is
not the case, the power expansion of the effective theory is
qualitatively different from the usual perturbation in $\alpha_s$,
and therefore worth exploring.

\section{Example: (1+1) Dimensions and Beyond}

We begin with an example that illustrates the basic ideas in a
simple setting. Consider the Euclidean (1+1) action of
non-relativistic fermions interacting with a gauge field A
\begin{equation}
\label{NRA} S = \int d\tau dx ~\psi^*_\sigma \left[ ( -
\partial_\tau + i\phi + \epsilon_F )  ~-~ \epsilon( -i\partial_x +
A ) \right] \psi_\sigma
\end{equation}
where $\epsilon (p)$ is the energy as a function of momentum (e.g.
$\epsilon (p) \approx {\frac{p^2}{2m}} + \cdots$). In (\ref{NRA})
and below we may consider it as a function of the operator
($-i\partial_x + A$). The dispersion relation in the presence of
the chemical potential $\epsilon_F$ is: $E(p) = \epsilon(p) -
\epsilon_F$, and a low energy mode must have momentum close to
$\pm p_F$, where $\epsilon ( \pm p_F ) = \epsilon_F$. The Fermi
surface in (1+1) dimensions is reduced to the two points $p = \pm
p_F$. Near these points we have
\begin{equation}
E( p \pm p_F ) \approx \pm~ v_F p~~~,
\end{equation}
where $v_F = \partial E / \partial p \vert_{p_F}$ is the Fermi
velocity.

The action (\ref{NRA}) is not obviously positive. In fact, the
operator in brackets $\left[ ~\cdots ~\right]$ clearly has
Hermitian as well as anti-Hermitian components, and hence complex
eigenvalues.

Let us assume that the gauge field has small amplitude and is
slowly varying relative to the scale $p_F$. We will extract the
slowly varying component of the fermion field to construct a low
energy effective theory involving quasiparticles and gauge fields.
This effective theory will have positive, semi-definite
determinant.

First, we extract the quasiparticle modes (we suppress the spin
index in what follows)
\begin{equation}
\psi (x, \tau) = \psi_L e^{+i p_F x} ~+~\psi_R e^{-i p_F x}~~~,
\end{equation}
where the functions $\psi_{L,R}$ are slowly varying. To simplify
the action, we use the identity
\begin{equation}
e^{\pm ip_F x} ~E( -i\partial_x + A )~ e^{\mp ip_F x} ~\psi (x)
\approx \pm~ v_F (-i\partial_x + A) \psi(x)~~~,
\end{equation}
to obtain\footnote{For simplicity we set $v_F = 1$. Alternatively it could be
absorbed in the definition of the spatial $\gamma_1$ as in QCD
below.}

\begin{equation}
\label{EFA} S_{\rm eff} = \int d\tau dx\left[\psi^{\dagger}_L ( -
\partial_\tau + i\phi + i\partial_x - A ) \psi_L  ~+~ \psi^*_R ( -
\partial_\tau + i\phi - i\partial_x + A ) \psi_R\right].
\end{equation}
We can write this in a more familiar form by introducing the
Euclidean (1+1) gamma matrices $\gamma_{0,1,2}$~, which are
Hermitian and can be taken as $\gamma_i = \sigma_i$ where
$\vec{\sigma}$ are the Pauli matrices. Using $\psi_{L,R} =
\frac{1}{2} ( 1 \pm \gamma_2) \psi$ we obtain
\begin{equation}
\label{SReff} S_{\rm eff} = \int d\tau dx ~  \bar{\psi}
\gamma^{\mu} (\partial_{\mu} + iA_{\mu} ) \psi ~\equiv~ \int d\tau
dx ~ \bar{\psi} D\!\!\!\!/ \psi ~~~.
\end{equation}
Since the gamma matrices are Hermitian, and the operator
$(\partial_{\mu} + iA_{\mu} )$ is anti-Hermitian, the operator
$D\!\!\!\!/$ in (\ref{SReff}) has purely imaginary eigenvalues.
However, because $\gamma_2$ anticommutes with $D\!\!\!\!/$, the
eigenvalues come in conjugate pairs: given $D\!\!\!\!/~ \phi =
\lambda \phi$, we have
$$D\!\!\!\!/ ~(\gamma_2 \phi) = - \gamma_2 D\!\!\!\!/ ~\phi = -
\gamma_2 \lambda \phi = - \lambda (\gamma_2 \phi_n)~~~.$$ Hence
the determinant $\det D\!\!\!\!/ ~= \prod \lambda^* \lambda$ is
real and positive semi-definite.

Thus, by considering only the low-energy modes near the Fermi
points of the original model (\ref{NRA}), we obtain an effective
theory with desirable positivity properties. Note that it is
necessary that the interactions (in this case, the background
gauge field A) not couple strongly the low-energy modes to fast
modes which are far from the Fermi points. This is a reasonable
approximation in many physical situations, where it is the
interactions among quasiparticles that are of primary interest. In
what follows, we will apply this basic idea to more complex models
such as QCD.

It is straightforward to go beyond (1+1) dimensions. Consider an
electron system, described by
\begin{equation}
{L}=\psi^{\dagger}\left[i\partial_t-{\epsilon(\vec p)}\right]\psi
+\mu\psi^{\dagger}\psi,
\end{equation}
where  $\epsilon(\vec p)$ is the electron energy, a function of
momentum $\vec p$. It is interesting to note that the
non-relativistic system already has a sign problem even at the
zero density, $\mu=0$, though the free case does not suffer this,
thanks to  the separation of variables. In fact, it is quite
unusual to have a system like vacuum QCD which has no sign
problem. In Euclidean space the electron determinant is
\begin{equation}
\label{edet} M=-\partial_{\tau}-\epsilon (\vec p)+\mu.
\end{equation}
The first term in operator (\ref{edet}) is anti-Hermitian, while
the rest are Hermitian. Since there is no constant matrix $P$ in
the spin space that satisfies $M^{\dagger}=P \, M\,P^{-1}$, it has
a sign problem in general.

Let us decompose the fermion momentum as
\begin{equation}
\vec p=\vec p_F+\vec l.
\end{equation}
Again, the Fermi momentum is defined to be a momentum at which the
energy equals to the chemical potential at zero temperature:
$\mu=\epsilon(p_F)$, and the Fermi velocity is defined as
\begin{eqnarray}
\vec v_F=\left.\frac{\partial\epsilon(p)}{\partial{\vec
p}}\right|_{p=p_F}.
\end{eqnarray}
If we are interested in low energies, $\left|\vec l\right|\ll
p_F$, we may integrate out the fast modes to get an effective
operator,
\begin{equation}
M_{\rm EFT}=-\partial_{\tau}-\vec v_F\cdot \vec l,
\end{equation}
which has  complex eigenvalues. However, when we include the
$-\vec v_F$ sector, we have $M_{\rm EFT}(\vec v_F)M_{\rm
EFT}(-\vec v_F)\le 0$ (i.e., has real negative eigenvalues),
assuming $\epsilon(\vec p)=\epsilon(-\vec p)$. We again see that
the sign problem in the electron system is alleviated in the
low-energy effective theory.

\section{QCD}

Let us recall why the measure of dense QCD is complex in Euclidean
space. We use the following analytic continuation of the Dirac
Lagrangian to Euclidean space:
\begin{equation}
 x_0 \rightarrow -i x_E^4,\quad x_i \rightarrow x_E^i
~;~ \gamma_0 \rightarrow \gamma_E^4,\quad \gamma_i \rightarrow
i\gamma_E^i ~~~.
\end{equation}
The Euclidean gamma matrices satisfy
\begin{equation}
{\gamma_E^{\mu}}^{\dagger}=\gamma_E^{\mu} ~~,~~
\left\{\gamma_E^{\mu},\gamma_E^{\nu}\right\}=2\delta^{\mu\nu}.
\end{equation}
The Dirac-conjugated field, $\bar\psi=\psi^{\dagger}\gamma^0$, is
mapped into a field, still denoted as $\bar\psi$, which is
independent of $\psi$ and transforms as $\psi^{\dagger}$ under
$SO(4)$. Then, the grand canonical partition function for QCD is
\begin{eqnarray}
Z(\mu)=\int {\rm d}A_{\mu}\det \left(M\right)e^{-S(A_{\mu})},
\end{eqnarray}
where $S(A_{\mu})$ is the positive semi-definite gauge action, and
the Dirac operator
\begin{equation}
\label{M} M=\gamma_E^{\mu}D_E^{\mu}+\mu\gamma_E^4,
\end{equation}
where $D_E = \partial_E + iA_E$ is the analytic continuation of
the covariant derivative. The Hermitian conjugate of the Dirac
operator is
\begin{equation}
M^{\dagger}=-\gamma_E^{\mu}D_E^{\mu}+\mu\gamma_E^4~~~.
\end{equation}
The first term in (\ref{M}) is anti-Hermitian, while the second is
Hermitian, hence the generally complex eigenvalues. When $\mu =
0$, the eigenvalues are purely imaginary, but come in conjugate
pairs $(\lambda, \lambda^*)$~\footnote{As before, note that $\gamma_5$
anti-commutes with M, so if $M \phi = \lambda \phi$, then $M
\gamma_5 \phi = - \gamma_5 M \phi = - \lambda \gamma_5 \phi$.},
so the resulting
determinant is real and positive semi-definite:
\begin{equation}
\rm det~ M =\prod \lambda^*\lambda \ge0~~~.
\end{equation}

In what follows we investigate the positivity properties of an
effective theory describing only modes near the Fermi surface. To
low energy modes, the curvature of the Fermi surface is not
evident, and the positivity of the usual Dirac sea determinant is
recovered.

A system of degenerate quarks with a net baryon number asymmetry
is described by the QCD Lagrangian density with a chemical
potential $\mu$,
\begin{equation}
{  L}_{\rm QCD}=\bar\psi i D\!\!\!\!/ ~\psi
-\frac{1}{4}F_{\mu\nu}^aF^{a\mu\nu}+\mu \bar\psi\gamma_0\psi,
\label{lag}
\end{equation}
where the covariant derivative $D_{\mu}=\partial_{\mu}+i A_{\mu}$
and we neglect the quark mass for simplicity.

At high density ($\mu\gg\Lambda_{\rm QCD}$), due to asymptotic
freedom the energy spectrum of quarks near the Fermi surface is
approximately given  by a free Dirac eigenvalue equation,
\begin{equation}
\left(\vec\alpha\cdot \vec
p-\mu\right)\psi_{\pm}=E_{\pm}\psi_{\pm},
\end{equation}
where $\vec\alpha=\gamma_0\vec\gamma$ and $\psi_{\pm}$ denote the
energy eigenfunctions with eigenvalues $E_{\pm}=-\mu\pm \left|\vec
p\right|$, respectively. At low energy $E<\mu$, the states
$\psi_+$ near the Fermi surface,  $|\vec p|\sim\mu$, are easily
excited but $\psi_-$, which correspond to the states in the Dirac
sea, are completely decoupled due to the presence of the energy
gap $\mu$ provided by the Fermi sea. Therefore the appropriate
degrees of freedom at low energy consist of gluons and $\psi_+$
only.

Now, we wish to construct an effective theory describing the
dynamics of $\psi_+$ by integrating out modes whose energy is
greater than $\mu$. Consider a quark near the Fermi surface, whose
momentum is close to $\mu\vec v_F$. Without loss of generality, we
may decompose the momentum of a quark into a Fermi momentum and a
residual momentum as
\begin{equation}
\label{decomp} p_{\mu}=\mu v_{\mu}+l_{\mu},
\end{equation}
where $v^{\mu}=(0,\vec v_F)$. Since the quark energy is given as
\begin{equation}
E=-\mu+\sqrt{(l_{\parallel}+\mu)^2+l_{\perp}^2},
\end{equation}
the residual momentum should satisfy
$(l_{\parallel}+\mu)^2+l_{\perp}^2\le4\mu^2$ with $\vec
l_{\parallel}=\vec v_F\vec l\cdot \vec v_F$ and $\vec
l_{\perp}=\vec l-\vec l_{\parallel}$.

To describe the small excitations of the quark with Fermi
momentum, $\mu\vec v_F$, we decompose the quark fields as
\begin{equation}
\label{psidecomp} \psi(x)=e^{i\mu\vec v_F\cdot \vec x}\left[
\psi_+(\vec v_F,x)+ \psi_-(\vec v_F,x)\right],
\end{equation}
where
\begin{equation}
\psi_{\pm}(\vec v_F,x)=P_{\pm}(\vec v_F) e^{-i\mu\vec v_F\cdot\vec
x}\psi(x) \quad{\rm with}\quad P_{\pm}(\vec v_F)\equiv\frac{1\pm
\vec \alpha\cdot\vec v_F}{2}.
\end{equation}
The quark Lagrangian in Eq.~(\ref{lag}) then becomes
\begin{eqnarray}
\label{expand} \bar\psi \left(i D\!\!\!\!/ +\mu\gamma^0\right)\psi
&=&  \bar\psi_+(\vec
v_F,x)i\gamma^{\mu}_{\parallel}D_{\mu}\psi_+(\vec v_F,x) \nonumber
\\ &+& \bar\psi_-(\vec v_F,x)\gamma^{0}\left(2\mu+i\bar
D_{\parallel}\right)\psi_-(\vec v_F,x) \nonumber \\ &+& \left[
\bar\psi_-(\vec v_F,x)i {D\!\!\!\!/}_{\perp}\psi_+(\vec
v_F,x)+{\rm h.c.} \right]
\end{eqnarray}
where $\gamma^{\mu}_{\parallel}\equiv(\gamma^0,\vec v_F\vec
v_F\cdot\vec \gamma)$, $\gamma^{\mu}_{\perp}=\gamma^{\mu}-
\gamma^{\mu}_{\parallel}$, $\bar D_{\parallel}=\bar
V^{\mu}D_{\mu}$ with $V^{\mu}=(1,\vec v_F)$, $\bar
V^{\mu}=(1,-\vec v_F)$, and ${D\!\!\!\!/}
_{\perp}=\gamma^{\mu}_{\perp}D_{\mu}$.

At low energy, we integrate out all the ``fast'' modes $\psi_-$
and derive the low energy effective Lagrangian by matching all the
one-light-particle irreducible amplitudes containing gluons and
$\psi_+$ in loop expansion. The effects of fast modes will appear
in the quantum corrections to the couplings of low energy
interactions. At tree-level, the matching is equivalent to
eliminating $\psi_-$ in terms of equations of motion:
\begin{equation}
\label{eliminate} \psi_-(\vec v_F,x)=-\frac{i\gamma^0}{2\mu
+iD_{\parallel}}{D\!\!\!\!/}_{\perp}\psi_+(\vec v_F,x)=-
\frac{i\gamma^0}{2\mu}
\sum_{n=0}^{\infty}\left(-\frac{iD_{\parallel}}{2\mu}\right)^n
{D\!\!\!\!/}_{\perp}\psi_+(\vec v_F,x). 
\end{equation}
Therefore, the tree-level Lagrangian for $\psi_+$ becomes
\begin{equation}
\label{treeL} {  L}_{\rm eff}^0=
\bar\psi_+i\gamma_{\parallel}^{\mu}D_{\mu}\psi_+-\frac{1}{2\mu}\bar\psi_+
\gamma^0({D\!\!\!\!/}_{\perp})^2\psi_+ ~+~ \cdots,
\end{equation}
where the ellipsis denotes terms with higher derivatives.

Consider the first term in our effective Lagrangian, which when
continued to Euclidean space yields the operator
\begin{equation}
M_{\rm eft}=\gamma^{E}_{\parallel}\cdot D(A).
\end{equation}
$M_{\rm eft}$ is anti-Hermitian and it anti-commutes with
$\gamma_5$, so it leads to a positive semi-definite determinant.
However, note that the Dirac operator is not well defined in the
space of $\psi_+(\vec v_F,x)$ (for fixed $v_F$), since it maps
$\psi_+(\vec v_F,x)$ into $\psi_+(-\vec v_F,x)$:
\begin{equation}
i
D_{\parallel}\!\!\!\!\!\!/~~P_+\psi=P_-iD_{\parallel}\!\!\!\!\!\!/~~\psi.
\end{equation}
Since $P_{-}(\vec v_F)=P_{+}(-\vec v_F)$, $iD\!\!\!\!/~\psi_+(\vec
v_F,x)$ are $\psi_+(-\vec v_F,x)$ modes, or fluctuations of a
quark with momentum $-\mu \vec v_F$.

Thus far we have considered the quark velocity as a parameter
labelling different sectors of the quark field. This is similar to
the approach of heavy quark effective theory
(HQET)\cite{Isgur:vq}, in which the velocity of the heavy charm or
bottom quark is almost conserved due to the hierarchy of scales
between the heavy quark mass and the QCD scale. However, this
approach contains an ambiguity often referred to as
``reparameterization invariance'', related to the non-uniqueness
of the decomposition (\ref{decomp}) of quark momenta into a large
and residual component. In the dense QCD case, two $\psi (v_F, x)$
modes whose values of $v_F$ are not very different may actually
represent the same degrees of freedom of the original quark field.
In what follows we give a different formulation which describes
{\it all} velocity modes of the quark field, and is suitable for
defining the quasiparticle determinant.

First, a more precise definition of the breakup of the quark field
into Fermi surface modes. Using the momentum operator in a
position eigenstate basis: $\vec{p} = -i \vec{\partial}$, we
construct the Fermi velocity operator:
\begin{equation}
\label{velo} \vec{v} =   \frac{-i }{\sqrt{- \nabla^2}}
~\frac{\partial}{\partial \vec{x}}~~,
\end{equation}
which is Hermitian, and a unit vector.

Using the velocity operator, we define the projection operators
$P_\pm$ as before and break up the quark field as, $\psi(x) =
\psi_+ (x) + \psi_- (x)$, with $\psi_\pm = P_\pm \psi$. By leaving
$\vec{v}$ as an operator we can work in coordinate space without
introducing the HQET-inspired velocity Fourier transform which
introduces $v_F$ as a parameter. If we expand the quark field in
the eigenstates of the velocity operators, we recover the previous
formalism with all Fermi velocities summed up.

The leading low-energy part of the quark action is given by
\begin{equation}
\label{leading} {  L}_+ =  \bar{\psi} P_- (v) \left( i
\partial\!\!\!/ - A\!\!\!/ + \mu \gamma_0
\right) P_+ (v) \psi~~.
\end{equation}
As before, we define the fields $\psi_+$ to absorb the large Fermi
momentum:
\begin{equation}
\label{absorb}
 \psi_+ (x) = e^{- i \mu \vec{x} \cdot \vec{v} } P_+
(v) \psi(x).
\end{equation}
Let us denote the eigenvalue $v$ obtained by acting on the field
$\psi$ (which has momentum of order $\mu$) as $v_l$ (or $v$
``large''), whereas eigenvalues obtained by acting on the
effective field theory modes $\psi_+$ are denoted $v_r$ (or $v$
``residual''). If the original quark mode had momentum p with $|p|
> \mu$ (i.e. was a particle), then $v_l$ and $v_r$ are parallel,
whereas if $|p| < \mu$ (as for a hole) then $v_r$ and $v_l$ are
anti-parallel. In the first case, we have $P_+ ( v_l ) = P_+
(v_r)$ whereas in the second case $P_+ (v_l) = P_- (v_r)$. Thus,
the residual modes $\psi_+$ can satisfy either of $P_\pm (v_r)
\psi_+ = \psi_+$, depending on whether the original $\psi$ mode
from which it was derived was a particle or a hole. In fact,
$\psi_+$ modes can also satisfy either of $P_\pm (v_l) \psi_+ =
\psi_+$ since they can originate from $\psi$ modes with momentum
$\sim + \mu v$ as well as $- \mu v$ (both are present in the
original measure: $D \bar{\psi} \, D \psi$). So, the functional
measure for $\psi_+$ modes contains all possible spinor functions
-- the only restriction is on the momenta: $|l_0|, |\vec{l}| <
\Lambda$, where $\Lambda$ is the cutoff.

In light of the ambiguity between $v_l$ and $v_r$, the equation
$\psi = e^{+ i \mu x \cdot v } \psi_+$ must be modified to
\begin{equation}
\psi = \exp \left( + i \mu x \cdot v ~ \alpha \cdot v \right)
\psi_+ = \exp \left( + i \mu x \cdot v_r ~\alpha \cdot v_r \right)
\psi_+ ~~~,
\end{equation}
where the factor of $\alpha \cdot v_r$ corrects the sign in the
momentum shift if $v_r$ and $v_l$ are anti-parallel. In general,
any expression with two powers of $v$ is unaffected by this
ambiguity. For notational simplicity we define a local operator
\begin{equation}
X ~\equiv~ \mu ~x\cdot v ~ \alpha \cdot v ~=~ \mu\frac{\alpha^i
x^j}{ \nabla^2}\frac{\partial^2}{\partial x^i
\partial x^j}.
\end{equation}

Taking this into account, we obtain the following action:
\begin{equation}
\label{leading1} {  L}_+ = \bar{\psi}_+ e^{- i X} \left( i
\partial\!\!\!/ - A\!\!\!/ + \mu \gamma_0
\right) e^{+ i X} \psi_+ ~~.
\end{equation}
We treat the $A\!\!\!/$ term separately from $i
\partial\!\!\!/ + \mu \gamma_0$ since the former does not commute
with X, while the latter does. Continuing to Euclidean space, and
using the identity $P_- \gamma_\mu P_+ = \gamma_\mu^\parallel
P_+$, we obtain
\begin{equation}
\label{leading2} {  L}_+ =  \bar{\psi}_+  \gamma^\mu_\parallel
\left(
\partial^\mu + i A^\mu_+ \right) \psi_+ ~~,
\end{equation}
where
\begin{equation}
A^\mu_+ = e^{-iX} ~ A^\mu ~e^{+iX}~~~,
\end{equation}
and all $\gamma$ matrices are Euclidean. The term containing $A$
cannot be fully simplified because $[v,A] \neq 0$. Physically,
this is because the gauge field carries momentum and can deflect
the quark velocity. The redefined $\psi_+$ modes are functions
only of the residual momenta l, and the exponential factors in the
A term reflect the fact that the gluon originally couples to the
quark field $\psi$, not the residual mode $\psi_+$.

The kinetic term in (\ref{leading2}) can be simplified to
\begin{equation}
\gamma^\mu_{\parallel} \partial^\mu = \gamma^\mu \partial^\mu
\end{equation}
since $v \cdot \partial \, v \cdot \gamma = \partial \cdot
\gamma~.$ The action (\ref{leading2}) is the most general
dimension 4 term with the rotational, gauge invariance\footnote{If
we simultaneously gauge transform $A_+$ and $\psi_+$ in
(\ref{leading2}) the result is invariant. There is a simple
relation between the gauge transform of the + fields and that of
the original fields: $U_+ (x) = U(x) e^{iX}.$ Of course, the
momentum-space support of the + gauge transform must be limited to
modes less than the cutoff $\Lambda$.} and projection properties
appropriate to quark quasiparticles. Therefore, it is a general
consequence of any Fermi liquid description of quark-like
excitations.

The operator in (\ref{leading2}) is anti-Hermitian and leads to a
positive, semi-definite determinant since it anti-commutes with
$\gamma_5$. The corrections given in (\ref{treeL}) are all
Hermitian, so higher orders in the $1/ \mu$ expansion may
re-introduce complexity. The structure of the leading term plus
corrections is anti-Hermitian plus Hermitian, just as in the
original QCD Dirac Lagrangian with chemical potential.

By integrating out the fast modes, the Euclidean QCD partition
function can be rewritten as
\begin{eqnarray}
Z(\mu)=\int {\rm d}A_+~\det M_{\rm eff}(A_+)e^{-S_{\rm eff}(A_+)}.
\end{eqnarray}
The leading terms in the effective action for gluons (these terms
are generated when we match our effective theory, with energy
cutoff $\Lambda$, to QCD) also contribute only real, positive
terms to the partition function:
\begin{equation}
S_{\rm eff}(A)=\int{\rm
d}^4x_E\left(\frac{1}{4}F_{\mu\nu}^aF_{\mu\nu}^a +\frac{M^2}{
16\pi}\sum_{\,\vec v_F}A_{\perp\mu}^{a}A_{\perp\mu}^{a}\right)
\ge0,
\end{equation}
where $A_{\perp}=A-A_{\parallel}$ and the Debye screening mass is
$M=\sqrt{N_f/(2\pi^2)}g_s\mu$\,. Note that Landau damping is due
to softer quark modes which have not been integrated out, and
therefore do not contribute to matching.

Although the HDET only describes low-energy modes, it still
contains Cooper pairing interactions. This is because Cooper
pairing, in which the quasiparticles have nearly equal and
opposite momenta, is induced by gluonic interactions with small
energy and momentum transfer. That is, although a gluon exchange
(or other interaction) which causes a large angular deflection of
a quasiparticle
$$ \vert \vec{p} \rangle \rightarrow \vert \vec{p}\,'\rangle $$
must involve a large momentum transfer, and hence is not part of
the effective theory, a Cooper pairing interaction
$$
 \vert \vec{p}, -\vec{p} \rangle \rightarrow \vert \vec{p}\,',- \vec{p}\,' \rangle
$$
only involves a small energy and momentum transfer, even if the
angle between $\vec{p}$ and $\vec{p}\,'$ is large. Hence, it is
described by the leading order interaction between soft gluons and
quarks in the effective theory (\ref{leading2}).

\section{Lattice Simulation}

The goal of this section is to give a method for simulating QCD at
finite density. We will consider a chemical potential $\mu$ much
larger than $\Lambda_{\rm QCD}$ throughout, and divide the
functional integral over quark excitations into two parts: (I)
modes within a shell of width $\Lambda$ of the Fermi surface, and
(II) modes which are further than $\Lambda$ from the Fermi
surface. We will assume the hierarchy
\begin{equation}
\label{hierarchy} \mu >> \Lambda >> \Lambda_{\rm QCD}~~~.
\end{equation}

The quark determinant in region (I) is well approximated by the
determinant of the leading operator in high density effective
theory (HDET) as long as the first inequality in (\ref{hierarchy})
is satisfied. As discussed in the previous section, it is positive
and real.

Here we will show that the contributions to the effective action
for the gauge field from quark modes in region (II) are small and
vanish as the $\Lambda$ grows large compared to $\Lambda_{\rm
QCD}$.

First consider the theory in Minkowski space. The Dirac operator
is
\begin{equation}
M = i D\!\!\!\!/ + \mu \gamma_0~~~
\end{equation}
and the Dirac equation can be written as
\begin{equation}
i \partial_0 \psi = H \psi~~~
\end{equation}
with
\begin{equation}
H = i \alpha \cdot \partial - \mu~~~
\end{equation}
a Hermitian operator. The break up into regions (I) and (II)
proceeds naturally in terms of energy eigenvalues of H (or $l_0$
in the HDET notation). The low-lying modes in region (I) are
particle states with spatial momenta satisfying $| \vec{p} |
\approx \mu$.

The analytic continuation of region (I) to Euclidean space leads
to the HDET determinant considered previously.

Modes in region (II) all have large energy eigenvalues, at least
as large as $\Lambda$. In considering their effect on physics at
the scale $\Lambda_{\rm QCD}$, we can integrate them out in favor
of local operators suppressed by powers of $\Lambda_{\rm QCD} /
\Lambda$.

To make this concrete, consider the effective action for gauge
fields with field strengths $F_{\mu \nu}$ of order $\Lambda_{\rm
QCD}$. The quark contribution to this effective action is simply
the logarithm of the determinant we wish to compute. It can be
expanded diagrammatically in graphs with external gauge field
lines connected to a single quark loop. Restricting to region
(II), we require that the quark modes in the loop have large H
eigenvalues. Evaluating such graphs leads only to operators which
are local in the external fields $A_\mu (x)$.

The resulting renormalizable (dimension 4) operator is the finite
density equivalent of $F_{\mu \nu}^2$, except that due to the
breaking of Lorentz invariance it contains separate time- and
space-like components which represent the contribution of
high-energy modes to the renormalization of the coupling constant,
and Debye screening. These effects do not introduce a complex
component when continued to Euclidean space.

Higher dimension operators, which involve additional powers of
$F_{\mu \nu}$ or covariant derivatives $D_\mu$ are suppressed by
the scale $\Lambda$. These are presumably the source of complex
terms introduced to the effective action. However, due to the
$1/\Lambda$ suppression they are dominated by the contribution
from the low-lying modes in region (I), which is necessarily
non-local, but real.

The logarithm of the Euclidean quark determinant will have the
form:
\begin{equation}
\label{lndet} {\rm ln \, det \,} M ~\sim~~ {  O}(\mu^4) ~+~ ( {\rm
non-local, real}) ~+~ {  O}( \frac{1}{\Lambda})({\rm local,
complex})~~~,
\end{equation}
where the first term is the (real, constant) free energy of
non-interacting quarks, the second term is from the positive
determinant in region (I) and the last term is the suppressed,
complex contribution from region (II). Only the last two terms
depend on the gauge field $A_\mu (x)$

On the lattice, one can use the dominant dependence of ${\rm det}
M$ on the first and second terms to do importance sampling. In
order to keep the complex higher dimension operators (last term in
(\ref{lndet})) small, it is important that the gauge field
strengths are kept smaller than $\Lambda^2$. One can impose this
condition by using two different lattice spacings, $a_{g}$ for the
gluons and  $a_{\det}$ for the quarks, with  $a_{g} > a_{\det}$.
The determinant is calculated on the finer $a_{\det}$ lattice, and
is a function of plaquettes which are obtained by interpolation
from the plaquettes on the coarser $a_{g}$ lattice. Interpolation
can be defined in a natural way, since each lattice link variable
$U_{x \mu}$ is an element of the gauge group, and one can connect
any two points $g_1, g_2$ on the group manifold in a linear
fashion: $g(t) = g_1 + t (g_2 - g_1)~,~ 0 \leq t \leq 1$.

\section{Inequalities and Anomaly Matching}

Positivity of the measure allows for rigorous QCD inequalities at
asymptotic density. For example, inequalities among masses of
bound states can be obtained using bounds on bare quasiparticle
propagators. One subtlety that arises is that a quark mass term
does not lead to a quasiparticle gap (the mass term just shifts
the Fermi surface). Hence, for technical reasons the proof of
non-breaking of vector symmetries\cite{Vafa:1984xg} must be
modified. (Naive application of the Vafa-Witten theorem would
preclude the breaking of baryon number that is observed in the
color-flavor-locked (CFL) phase\cite{Alford:1998mk}). A
quasiparticle gap can be inserted by hand to regulate the bare
propagator, but it will explicitly violate baryon number. However,
following the logic of the Vafa-Witten proof, any symmetries which
are preserved by the regulator gap cannot be broken spontaneously.
One can, for example, still conclude that isospin symmetry is
never spontaneously broken (although see below for a related
subtlety). In the case of three flavors, one can introduce a
regulator $d$ with the color and flavor structure of the CFL gap
to show rigorously that none of the symmetries of the CFL phase
are broken at asymptotic density. On the other hand, by applying
anomaly matching conditions\cite{anomaly}, we can prove that the
$SU(3)_A$ symmetries {\it are} broken. We therefore conclude that
the CFL phase is the true ground state for three light flavors at
asymptotic density, a result that was first established by
explicit
calculation\cite{Evans:1999at,Hong:2000ru,Schafer:1999fe}.

To examine the long-distance behavior of the vector current, we
note that its correlator in a given background
gauge field $A$ can be written as
\begin{eqnarray}
\left<J_{\mu}^a(\vec v_F,x)J_{\nu}^b(\vec v_F,y)\right>^A =-{\rm
Tr}\,\gamma_{\mu}T^a S^A(x,y;d)\gamma_{\nu}T^b S^A(y,x;d),
\nonumber
\end{eqnarray}
where the $SU(N_f)$ flavor current $J_{\mu}^a(\vec v_F,x)
=\bar\psi_+(\vec v_F,x)\gamma_{\mu}T^a\psi_+(\vec v_F,x)$. The
propagator with $SU(3)_V$-invariant IR regulator $d$ is given as
\begin{eqnarray}
S^A(x,y;d)=\left<x\right|\frac{1}{M}\left|y\right>=\int_0^{\infty}
{\rm d}\tau \left<x\right|e^{-i\tau (-iM)}\left|y\right> \nonumber
\end{eqnarray}
where with $D=\partial+iA$
\begin{eqnarray}
M &=& \gamma_0
\begin{pmatrix}
D\cdot V \hfill & d \\
d^{\dagger} & D\cdot\bar V \hfill
\end{pmatrix}  \nonumber \\
\end{eqnarray}
Since the eigenvalues of $M$ are bounded from below by $d$, we
have
\begin{eqnarray}
\left|\left<x\right|\frac{1}{M}\left|y\right>\right| \le
\int_R^{\infty}\!\!\!{\rm d}\tau \,e^{-d \,
\tau}\sqrt{\left<x|x\right>} \sqrt{\left<y|y\right>}=\frac{e^{-d
\, R}}{d} \sqrt{\left<x|x\right>}\sqrt{\left<y|y\right>},
\label{propagator1}
\end{eqnarray}
where $R\equiv\left|x-y\right|$. The current correlators fall off
rapidly as $R\to \infty$;
\begin{eqnarray}
\left| \int {\rm d}A_+\!\!\!\right.\!\!\!\!\!& &\left.\!\!\!\!\! ~
\det M_{\rm eff}(A)\,\,e^{-S_{\rm eff}}
\left<J_{\mu}^A(\vec v_F,x)J_{\nu}^B(\vec v_F,y)\right>^{A_+}\right|\nonumber\\
&\le & \int_{A_+} \left| \left<J_{\mu}^A(\vec v_F,x)J_{\nu}^B(\vec
v_F,y) \right>^{A_+}\right| \le \frac{e^{-2d \, R}}{d^2}
\int_{A_+}
\left|\left<x|x\right>\right|\left|\left<y|y\right>\right|,
\label{schwartz}
\end{eqnarray}
where we used the Schwartz inequality in the first inequality,
since the measure of the effective theory is now positive, and
equation (\ref{propagator1}) in the second inequality. The IR
regulated vector currents do not create massless modes out of the
vacuum or Fermi sea, which implies that there is no
Nambu-Goldstone mode in the $SU(3)_V$ channel. Therefore, for
three massless flavors $SU(3)_V$ has to be unbroken as in CFL. The
rigorous result provides a non-trivial check on explicit
calculations, and applies to any system in which the quasiparticle
dynamics have positive measure. The case with non-zero quark
masses is complicated, and requires careful consideration of the
order of limits\cite{Hong:2002nn}.

\section{Conclusion}

The low-energy physics of dense fermionic matter, ranging from
quark matter to electronic systems, is controlled by modes near
the Fermi surface. An effective Lagrangian describing the
low-energy modes can be given in a systematic expansion in powers
of the energy scale over the chemical potential. The leading term
in this expansion has a simple form, and we have shown that it
leads to a real, positive Euclidean path integral measure.

This observation opens the door to importance sampling in Monte
Carlo simulations of dense matter systems. The key requirement is
that the interactions do not strongly couple the low-energy modes
to modes far from the Fermi surface. QCD at high density satisfies
this requirement, as do all asymptotically free models. Electronic
systems in which the important interactions involve momentum
transfer less than the Fermi energy are in this category, although
some idealized models such as the Hubbard model are not. We have
given some proposals for how the positive effective theory might
be simulated numerically. Ultimately, we hope that actual
practitioners will develop even more practical methods.

Finally, positivity has analytical applications as well, since it
allows the use of rigorous inequalities. In QCD we obtain
restrictions on symmetry breaking patterns at high density.
Similar restrictions can probably be obtained for electronic
systems with suitable interactions.

\end{document}